\documentclass{bvm2022} 

\begin{document}



\selectlanguage{english} 

\title{Effect of Random Histogram Equalization on Breast Calcification Analysis Using Deep Learning}

\titlerunning{Effect of RHE on Breast Calcification Analysis Using DL}

\author{
	Adarsh\lname{Bhandary~Panambur}\inst{1,2}, 
	Prathmesh\lname{Madhu}\inst{2}, 
	Andreas\lname{Maier}\inst{2} 
}

\authorrunning{Bhandary~Panambur, Madhu \& Maier}

\institute{
\inst{1} Technology and Innovation Management, Siemens Healthineers, Erlangen, Germany\\
\inst{2} Pattern Recognition Lab, FAU Erlangen-N\"urnberg, Erlangen, Germany\\
}

\email{adarsh.bhandary.panambur@fau.de}

\maketitle

\begin{abstract}

Early detection and analysis of calcifications in mammogram images is crucial in a breast cancer diagnosis workflow. Management of calcifications that require immediate follow-up and further analyzing its benignancy or malignancy can result in a better prognosis. Recent studies have shown that deep learning-based algorithms can learn robust representations to analyze suspicious calcifications in mammography. In this work, we demonstrate that randomly equalizing the histograms of calcification patches as a data augmentation technique can significantly improve the classification performance for analyzing suspicious calcifications. We validate our approach by using the CBIS-DDSM dataset for two classification tasks. The results on both the tasks show that the proposed methodology gains more than 1\% mean accuracy and F1-score when equalizing the data with a probability of 0.4 when compared to not using histogram equalization. This is further supported by the t-tests, where we obtain a \emph{p}-value of \emph{p}<0.0001, thus showing the statistical significance of our approach.  
\end{abstract}

\section{Introduction}
\label{3258-sec-introduction}

With an estimated 2.26 million new cases in 2020, breast cancer is the leading cause of malignancy incidence worldwide \cite{3258-01}. Periodic screening of women based on their age groups using mammography has proven to reduce the late-stage incidence and mortality rates \cite{3258-01}. Breast Imaging Reporting and Database System (BI-RADS) is a scoring system used by radiologists to assign categories to the suspicious region of interest on screening mammography based on the type, density, morphology, and distribution of the findings \cite{3258-02}. In addition to observing essential findings such as mass, density, architectural distortions, and asymmetries in the breast, the clinicians also search for bright well-circumscribed findings known as calcification. Calcifications are calcium deposits, generally found inside the ducts, lobules, connective stroma tissue, and vessels of the breast \cite{3258-03}. The analysis of calcification on mammograms is of high importance as its morphology and distribution can be representative of precancerous or malignant cells \cite{3258-03}. As the sensitivity of calcification detection during the initial mammography screening is low \cite{3258-04}, computer-aided diagnosis (CAD) methods can potentially support clinicians for better decision making for early detection and analysis of calcifications. Deep learning (DL)-based CAD algorithms have proven to provide effective and robust solutions for automated breast cancer analysis in mammography \cite{3258-05}, with many recent approaches showing promising results in calcification analysis tasks \cite{3258-06,3258-07}. 

Contrast enhancement techniques have been vastly studied as a pre-processing image technique in whole mammogram image analysis \cite{3258-08}. However, applying these techniques to smaller calcification patches might lead to a loss of class-specific information. This is evident due to the appearance of calcification which generally appears as a bright spot or as a group of scattered bright spots, and equalizing the histograms of such patches might lead to noisy images. Adding noise to the input data is one of the data augmentation techniques used for training DL networks \cite{3258-09}. Data augmentation is the most commonly used strategy in DL to increase the quality of the training data by applying transformations to the input enforcing the network to learn meaningful, class-specific representations \cite{3258-10}. In this work, we investigate the usage of random histogram equalization (RHE) as a data augmentation technique for calcification analysis. Histogram equalization (HE) is a contrast enhancement technique that non-linearly scales the input to a uniform distribution of intensities based on the image's histogram. 

The main contributions of our research are:
a) We present that using RHE as a data augmentation technique with a probability value of 0.4 can lead to a significant performance gain compared to not using RHE; 
b) We also present and compare Gradient-weighted Class Activation Mapping (Grad-CAM) \cite{3258-10} for models that were trained using no HE and full HE to show that RHE outperforms both of them in localizing class-specific calcification features.

The remaining organization of the paper is as follows: Section \ref{3258-sec-materialsmethods} provides a brief description of the data and methods used in our work. Next, we present the quantitative and qualitative results for two and three-class classification tasks in Section \ref{3258-sec-results}. Finally, in Section \ref{3258-sec-discussion}, we provide a brief discussion on the results.

\section{Materials and methods}
\label{3258-sec-materialsmethods}

\subsection{Data}
\label{3258-sec-data}

Curated Breast Imaging Subset of Digital Database for Screening Mammography (CBIS-DDSM) is a public dataset consisting of scanned film mammography from multiple institutions across the United States of America \cite{3258-11}. In this research, we utilize a small subset of the dataset consisting of cropped regions of calcifications observed on craniocaudal and/or mediolateral oblique views in the mammogram images.

\begin{table}[h]
\begin{tabular*}{\textwidth}{l@{\extracolsep\fill}ccccc}

\hline

         & Follow-up & No follow-up  & Malignant  & Benign  & Benign w/o callback\\  \hline
Training    & 855        & 382           & 445         & 410     & 382  \\
Validation  & 217        & 92            & 99          & 118     & 92  \\   
Test        & 259        & 67            & 129         & 130     & 67  \\       
Total       & 1331       & 541           & 673         & 658     & 541  \\ \hline
\end{tabular*}
\caption{The distribution of class samples in the CBIS-DDSM dataset for the two-class (Follow-up versus No follow-up) and three-class problems (Malignant, benign and benign without callback).}
\label{3258-tab1}
\end{table}

In Table~\ref{3258-tab1}, we show the distribution of the class samples in the training, validation, and test dataset. The dataset consists of 1,546 images divided into training and validation datasets with an 80:20 ratio. In addition, we use an independent test dataset from CBIS-DDSM with 326 images as the test dataset. The verified pathology information available in this dataset is used as the labels for the two classification tasks. The first task involves the binary classification of  \elqq Follow-up\erqq~versus \elqq No follow-up\erqq~calcification patches. The calcifications labeled as \elqq malignant\erqq~and \elqq benign\erqq~are used for the \elqq Follow-up\erqq~class as these patches would require additional clinical investigations. \elqq No follow-up\erqq~class consists of \elqq benign without callback\erqq~calcifications indicating the lesions are worth tracking but does not require further investigations. Finally, we use the original \elqq malignant\erqq~, \elqq benign\erqq~, and \elqq benign without callback\erqq~labels for the three-class classification.

\subsection{Experimental setup}
\label{3258-sec-expsetup}
Based on our initial experiments with various state-of-the-art convolutional neural networks (CNN), we chose a pre-trained ResNet50 \cite{3258-12} as the standard CNN architecture for all the experiments conducted in this research. The final fully connected layer (head) in the CNN is refactored to represent the number of classes, i.e., 2 and 3, respectively. The input images are resized with bilinear interpolation into 224 x 224 pixels, and the pixel values are normalized to the range \emph{[0,1]}. Four fixed data augmentations are used to enhance the quality and size of the training data. Random horizontal flip, random vertical flip, random rotations are used with a probability value of 0.5, and random erasing is used with a probability value of 0.1. The probability value indicates the possibility of the current image being transformed. As this research aims to investigate the effect of RHE as a data augmentation technique in calcification analysis, different experiments with probability values (\emph{P}) of 0, 0.2, 0.4, 0.6, 0.8, and 1 are conducted. All the CNNs are trained for 30 epochs with a batch size of 16. A weighted binary cross-entropy loss function is optimized using a standard Adam optimizer with a learning rate of $3.2e^{-6}$ and a weight decay of $1e^{-4}$.

Accuracy and F1-score are the two performance evaluation metrics used for the classification task. We further utilize a two-tailed unpaired t-test in order to determine the statistical significance of the classification performances. Finally, the qualitative performance of the CNN is analyzed by localizing the regions in the calcification patch that influenced the model's decision by using the Grad-CAM approach \cite{3258-10}. The entire DL pipeline was developed using PyTorch.

\section{Results}
\label{3258-sec-results}

\begin{table}[h]
\caption{Classification performance of the trained CNN on the test dataset. Different values of \emph{P} indicate the probability of RHE on a patch during training. The best results are in \emph{italics}.}
\label{3258-tab2}
\begin{tabular*}{\textwidth}{l@{\extracolsep\fill}cccc}

\hline                          
\multirow[t]{2}{*} & \multicolumn{2}{c}{Two-Class Task} & \multicolumn{2}{c}{Three-Class Task} \\
         & Accuracy                & F1-Score                    & Accuracy                  & F1-Score \\  \hline
\emph{P}= 0      & 0.9215 (0.0066)          & 0.8838 (0.0078)            &  0.6632 (0.0126)          & 0.6855 (0.0131)        \\
\emph{P}= 0.2    & 0.9159 (0.0117)          & 0.8759 (0.0117)            &  0.6755 (0.0191)          & 0.6985 (0.0221)        \\
\emph{P}= 0.4    & \emph{0.9325 (0.0085)} & \emph{0.8973 (0.0124)}   &  \emph{0.6840 (0.0193)} &\emph{0.7071 (0.0189)}\\
\emph{P}= 0.6    & 0.9196 (0.0092)          & 0.8770 (0.0119)            & 0.6687 (0.0058)           & 0.6969 (0.0036)        \\
\emph{P}= 0.8    & 0.9252 (0.0074)          & 0.8869 (0.0109)            & 0.6601 (0.015)            & 0.6839 (0.0159)        \\
\emph{P}= 1      & 0.8724 (0.0179)          & 0.8103 (0.0111)            & 0.6252  (0.017)           & 0.6322 (0.0228)        \\
\hline
\end{tabular*}
\end{table}

In Table~\ref{3258-tab2}, we summarize the performance of the trained CNNs on the test dataset for both the two-class and three-class classification tasks. The different values of \emph{P} in the results denote the probability of a calcification patch being equalized during the training  (Tab.~\ref{3258-tab2}). The probability value of 0 (\emph{P}=0) indicates that no HE is used during training or testing, whereas a value of 1 (\emph{P}=1) indicates all the images are equalized during both training and testing phases. For all the other intermediate values (\emph{P}=0.2, 0.4, 0.6, 0.8), RHE is used only during training. Each experiment for each probability value was run five times to ensure stability in the final reported results. The mean accuracy and F1-score with the standard deviation over five validation runs are reported on the test dataset. In Figure~\ref{3258-figure1}, the Grad-CAM visualizations comparing the network outputs for best-performing models for \emph{P}=0 and \emph{P}=0.4 are shown for both the tasks. Figure~\ref{3258-figure2} shows the Grad-CAM visualizations when all the input training calcification patches are histogram equalized.

\section{Discussion}
\label{3258-sec-discussion}

 As observed in the results summarized in Table~\ref{3258-tab2}, the network learns better representations for both the classification tasks using RHE during training with a probability of 0.4 compared to not using HE (\emph{P}=0). For the two-class task, i.e., Follow-up vs. No follow-up, we achieve a mean accuracy and F1-score of 0.9215 and 0.8838, respectively, on the test dataset, when no HE is applied during training. When we increase the probability to a value of 0.4, we reach a mean accuracy of 0.9325 and a mean F1-score of 0.8973. A performance boost of more than 1\% is observed in both the mean accuracy and F1-score. The results are then tested for statistical significance of accuracy and F1-score using a t-test, where a two-tailed \emph{p}-value of \emph{p}<0.0001 is obtained. This clearly shows that using RHE with a probability of 0.4 results in a statistically significant performance. A similar trend is observed in malignant, benign, and benign without callback classification. In this case, we achieve more than 2\% gain in both the metrics by using the probability value of 0.4, with mean accuracy increasing from 0.6632 to 0.6840 and the F1-score increasing from 0.6855 to 0.7071. A \emph{p}-value of \emph{p}<0.0001 is obtained, showing a statistically significant classification performance during the t-test.

\begin{figure}[h]
	\centering 
	\setlength{\figwidth}{0.49\textwidth}
	\caption{Input image with true class label, Grad-CAM image for \emph{P}=0 and \emph{P}=0.4 with the predicted class labels. The localized regions responsible for misclassifications can be observed while using \emph{P}=0. However, using \emph{P}=0.4, the network is able to learn robust class-discriminative features.}
	\label{3258-figure1}
	\begin{subfigure}{\figwidth}
		\includegraphics[width=\textwidth]{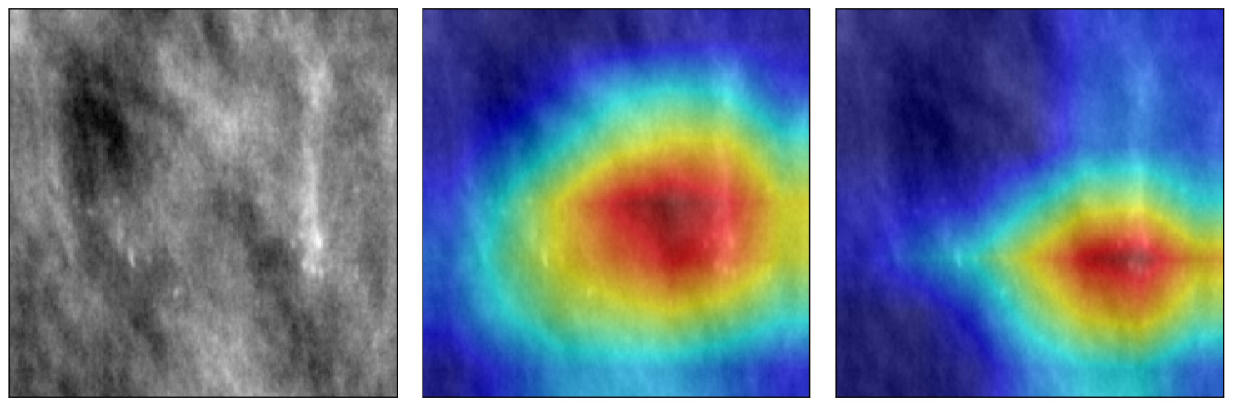}
		\caption{Malignant, Benign, Malignant.}
	\end{subfigure}
	\hfill	
	\begin{subfigure}{\figwidth}
		\includegraphics[width=\textwidth]{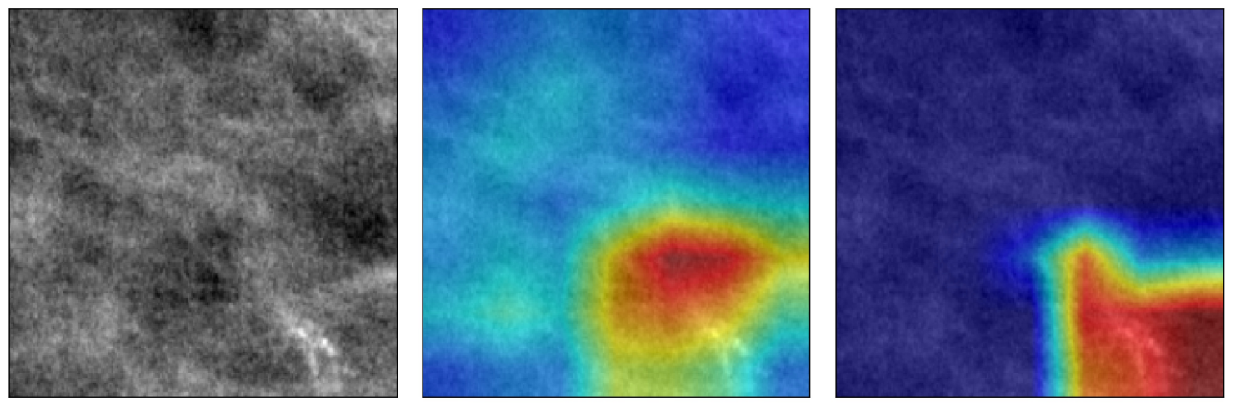}
		\caption{Follow-up, No follow-up, Follow-up.}
	\end{subfigure}
	\hfill	
\end{figure}

\begin{figure}[h]
	\centering 
	\setlength{\figwidth}{0.49\textwidth}
	\caption{Input image with true class label, histogram equalized image and the Grad-CAM image with the predicted class label for \emph{P}=1. The network is not able to extract the class-relevant features even though the it is able to localize the suspicious region of calcification.}
	\label{3258-figure2}
	\begin{subfigure}{\figwidth}
		\includegraphics[width=\textwidth]{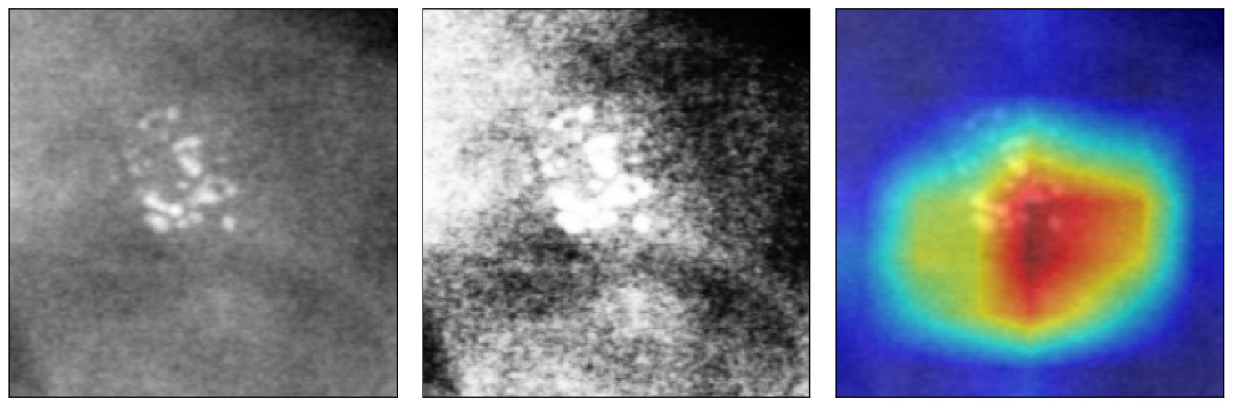}
		\caption{Malignant, Benign.}
	\end{subfigure}
	\hfill	
	\begin{subfigure}{\figwidth}
		\includegraphics[width=\textwidth]{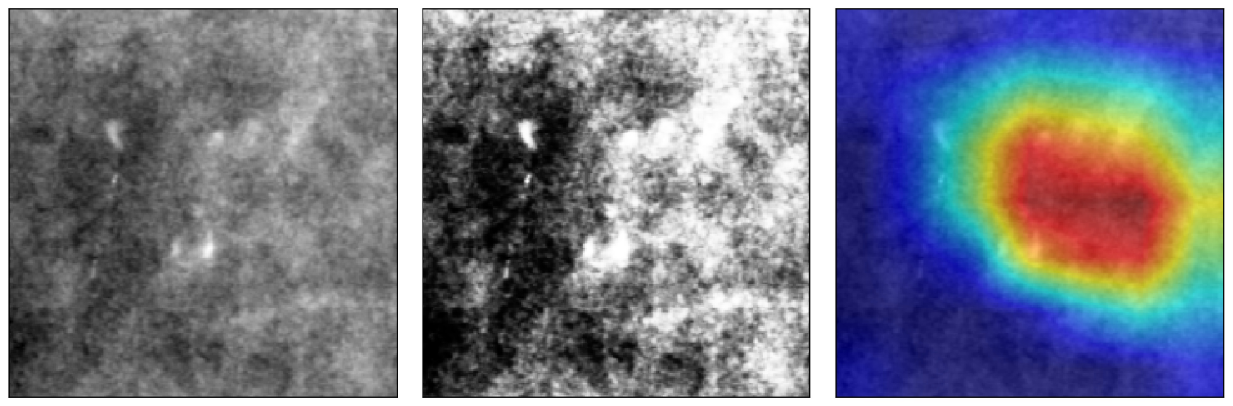}
		\caption{Follow-up, No follow-up.}
	\end{subfigure}
	\hfill	
\end{figure}

The increase in performance can be attributed to the randomly injected noise in the form of RHE during training, forcing the network to concentrate on the calcification regions rather than the surrounding anatomical structures. This behavior can be observed using Grad-CAM visualizations in Figure~\ref{3258-figure1}. The models trained with random probability of \emph{P}=0.4 are inherently more robust and tend to focus more towards the high-intensity spots indicative of calcifications. The results in Table~\ref{3258-tab2} also show a significant reduction in the performance while using \emph{P}=1, i.e., equalizing the histograms of all images during training and testing. This is due to the enhancement of the contrast in the brighter parts of the image leading to corruption of the class-specific features in the data. In Figure~\ref{3258-figure2}, we can observe that the essential features such as the morphology and the distribution of the calcifications are corrupted due to the equalization of the histogram values in the image. Even though the attention of the network is focused towards the region of calcifications, we observe misclassifications.

As the current work involves calcification patches detected on scanned film mammography, we plan to validate the proposed technique using digital mammography images as a part of future work. None of the research involving patch-level calcification analysis has exclusively explored the effect of data augmentations on model performance. We aim to perform additional ablation studies to examine the impact of other data augmentations techniques on calcification analysis.

\printbibliography

\end{document}